\documentclass[12pt]{article}
\usepackage{epsfig}
\usepackage{a4}

\begin{document}

\newcommand \be  {\begin{equation}}
\newcommand \bea {\begin{eqnarray} \nonumber }
\newcommand \ee  {\end{equation}}
\newcommand \eea {\end{eqnarray}}

\title{\bf Experts' earning forecasts: bias, herding and gossamer information}

\author{Olivier Guedj$^*$, Jean-Philippe Bouchaud$^*$}
\maketitle
{\small
$^*$ Science \& Finance, Capital Fund Management, 6-8 Bvd Haussmann\\
75 009 Paris, France}\\

\begin{abstract}
We study the statistics of earning forecasts of US, EU, UK and JP stocks 
during the period 1987-2004. We confirm, on this large data set,  
that financial analysts are on average over-optimistic and show a 
pronounced herding behavior. These effects are time 
dependent, and were particularly 
strong in the early nineties and during the Internet bubble. We furthermore 
find that their forecast ability is, 
in relative terms, quite poor and comparable in quality, a year ahead, to the simplest `no change' forecast. As a result of
herding, analysts agree with each other five to ten times more than with the actual result. We have shown that significant differences 
exist between US stocks and EU stocks, that may partly be explained as a company size effect. Interestingly, herding 
effects appear to be stronger in the US than in the Eurozone. Finally, we study the correlation of errors across 
stocks and show that significant sectorization occurs, some sectors being easier to predict than others. These results
add to the list of arguments suggesting that the tenets of Efficient Market Theory are untenable. 
\end{abstract}



\section{Introduction}

The Efficient Market Hypothesis (EMH) assumes that prices reflect faithfully all available 
information \cite{Fama,Samuelson}. Agents are rational and use all economic information to 
form an unbiased estimate of future dividends; the price is at any instant of time the best predictor of 
future prices. This hypothesis is of crucial importance for policy making and 
strategic corporate decisions, and also, of course, for investing in stock markets. Arguments against 
this idealized view of the world are however mounting. The very notion of `information' needs to be
clarified. Is there anything like reliable, unbiased information on which rational agents can base their
anticipations? In a complex system such as the economic world, much as in complex systems from the
physical world, tiny perturbations can cause major disturbances, and any small imprecision can lead to a
total lack of predictability \cite{Arthur,Ruelle,Bak}. Any information item has to be
processed and interpreted, and is at best a very noisy indicator of future results. If information 
is intrinsically noisy, then prices will also be noisy; this means that prices can wander freely within 
a noise band without being clearly absurd. This relates to Black's (somewhat provocative) idea that an efficient market 
is such that prices are within a factor two from their `true' values \cite{Black}. Within this `noise band', price
moves are mainly due to trading and speculation, based on gossamer information. 
This leads to a short to medium term volatility much too high to be compatible with the idea of 
fully rational pricing \cite{Shiller}. 

How reliable is the information available to market participants? A way to analyze this question 
is to study how well market experts forecast the future earnings of companies. In principle, all
publicly available information is known to them. Their academic education and professional training 
should give them all the necessary tools to extract from this information an optimal forecast of the
future earnings of the companies they specially focus on. Ideally, these forecasts should be unbiased and with a 
rather small prediction error. Since individuals are prone to errors, misinterpretations and personal biases,
a pool of specialists should improve the situation and collectively reduce any forecasting error. Actually,
financial markets are often seen as efficient because of the aggregation, through offer and demand, of 
a large number of differing opinions on the price; the final price therefore reflecting the collective
consensus of market participants. 

Interestingly, the statistical analysis of security analyst forecasts is possible using various databases, and
has already been the subject of several studies in the past (see e.g. \cite{Trueman,Forbes,Bagella,Hong,Krause}). 
The much debated findings are (i) the systematic upward bias of the analysts' predictions (over-optimism) and (ii) a herding tendency between analysts. As expressed
by B. Trueman, {\it ``analysts exhibit herding behavior, whereby they release forecasts similar to those previously 
announced by other analysts, even when this is not justified by their information''}\cite{Trueman}. There are many papers
in the literature discussing the reality and significance of these findings, their possible causes and their 
impact on the behavior of investors. The aim of this paper is to give the results
of an extensive analysis of available data concerning forecasts of US, European, UK and JP stocks earnings, over the
period 1987-2004. Our conclusions, summarized below, 
are in broad agreement with the results reported in previous papers, but our results allow us to make more precise 
statements and discuss 
some claims and conjectures made the literature. We also give results that, to our knowledge, have not been
discussed before, such as the cross-sectional correlations between forecast errors.

Our main results, that we will comment later in the paper, are: 
\begin{itemize}

\item (i) there is an overall positive bias in the analyst forecasts, that varies somewhat over the
years -- small in the mid nineties, substantial during the Internet bubble -- and systematically decreases as 
one approaches the earning announcement date ({\sc ead}). The relative bias is on average of $60 \%$ a year before 
{\sc ead} and decreases to $10 \%$ one month before {\sc ead}. 

\item (ii) the bias is stronger on stocks not belonging to the S\&P than on ones that belong to the S\&P, and stronger 
on EU, UK or JP liquid stocks than on S\&P stocks. Our results also clearly suggest that the bias is negatively correlated with
company size. 

\item (iii) the relative forecast error is on average rather large and in any case a factor 3-10 larger than the 
dispersion of forecasts among different analysts! This is a very strong hint of herding behavior. Both the
forecast error and the dispersion decrease as the {\sc ead} is approached. Interestingly, the herding behavior 
appears to be less pronounced in the EU than in the US, the UK or JP.

\item (iv) one year before the next earning announcement, the simple `no-change' forecast (i.e the 
next earning will equal the last one) slightly outperforms the analyst forecasts: it is on average
less biased, and has a similar forecast error.

\item (v) there is a non trivial structure in the correlation of prediction errors, that shows clear signs of 
sectorization. Some sectors appear to be easier to predict than others, while significant intra-sector correlations
between forecast errors can be observed.

\end{itemize}

As we discuss in the conclusion, these findings are hard to reconcile, both at the qualitative and 
quantitative level, with the idea of rational anticipations and efficient markets. 

\section{Presentation of the data}

The Institutional Brokers Estimate System provides investment professionals with a global database 
of analyst forecasts of earnings per share ({\sc eps}), cash flow per share, dividends per share, and net profit per share, 
plus additional measures such as sales, EBIT, EBITDA and recommendations for publicly traded corporations worldwide. 
Thomson Financial Datastream provides both analyst detailed forecasts and consensus forecasts over a large period of time.

The universe from which we extract our sample is represented by 1663 US companies, with 491  belonging to the S\&P, 
445 EU companies, 402 JP companies and 302 UK companies. Both real {\sc eps} and consensus forecasts come from 
Datastream database. On average, 10.66 analysts participate to the consensus on all US stocks, 16.36 on S\&P stocks, 
17.26 on EU stocks, 8.67 on JP stocks and 10.10 on UK ones. In order to reduce the noise due to outliers, both earnings per share and 
forecasts have been restricted to values in $[-10;10]$, over the period 1987-2004, where data are available for all 
studied pools.

\section{Forecast bias and forecast error}

\subsection{Definitions}

We will study below the annual {\sc eps} normalized by the price of a share, that we will call
$\varepsilon_{\alpha}$, where $\alpha$ labels the stocks. The announced earning at time $t$ is then noted 
$\varepsilon_{\alpha,t}$. The forecast made by expert $i$ 
of $\varepsilon_{\alpha,t}$ at time $t -\theta$ before the {\sc ead} is $f^i_{\alpha,t-\theta}$, while 
$f_{\alpha,t-\theta}$ is the average over experts of the same quantity. The dispersion (root mean square) of forecasts 
over analysts is $\sigma_{\alpha,t-\theta}$. As mentioned above, forecasts are released every months by a pool of $\approx 11$ analysts for
US stocks (averaged over stocks). The ex-post forecast bias is obviously defined as:
\be
b_{\alpha,t-\theta} = f_{\alpha,t-\theta} - \varepsilon_{\alpha,t}.
\ee
The grand average of the bias for a fixed distance $\theta$ from the {\sc ead}, over both stocks and years, is 
$B(\theta)$; one can also study how the average bias for a fixed $\theta$ depends on time by averaging over 
$\alpha$ for a fixed $t$; the corresponding quantity is $B(\theta;t)$. The forecast error $\Sigma$ is defined as
\footnote{Note that since the average bias $B$ turns out to be very small compared to the forecast error $\Sigma$,
it is irrelevant to subtract or not $B^2$ from the following definitions.}
\be
\Sigma^2(\theta)= \left \langle \left(f_{\alpha,t-\theta} - \varepsilon_{\alpha,t}\right)^2 \right \rangle_{\alpha,t},
\ee
and:
\be
\Sigma^2(\theta;t)= \left \langle \left(f_{\alpha,t-\theta} - \varepsilon_{\alpha,t}\right)^2 \right \rangle_{\alpha},
\ee
where $\langle .. \rangle_x$ denotes the average over $x$. Similarly, we will be interested by the grand average of the 
forecast dispersion over time and stocks, $\sigma(\theta)$, and the year dependent average forecast dispersion
over stocks, $\sigma(\theta;t)$. 

\subsection{Forecast bias}

\begin{figure}
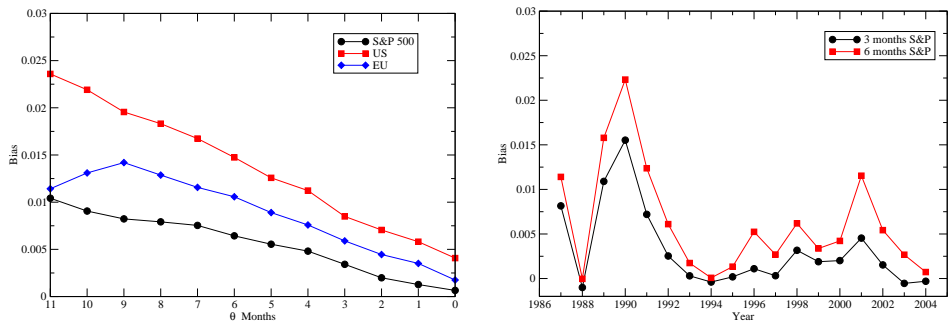

\begin{center}
\epsfig{file=bias.eps,width=6cm,clip=}\hspace{0.5cm}\epsfig{file=biasm.eps,width=6cm,clip=} 
\end{center}
\caption{Left: The grand average bias $B(\theta)$ for S\&P stocks, US stocks (including S\&P stocks) and European stocks, 
as a function of the time $\theta$ to {\sc ead}. The statistical error bar is of the order of the size of the
symbols. Note that the bias is smallest for S\&P stocks, intermediate for EU stocks, and largest for the US stock
pool, suggesting a size effect (see Fig. 2). Right: The bias $B(\theta;t)$ for S\&P stocks, for $\theta=3$ months and $\theta=6$ months,
as a function of the time $t$ (year of the earning announcement). Note that the bias was very small around 1994 and 2003.}
\label{Fig1}
\end{figure}

The grand average bias $B(\theta)$ is plotted as a function of $\theta$ in Fig. 1 (left), for three different ensemble of
stocks: S\&P stocks, large pool of US stocks and European stocks. We observe that the bias is positive,\footnote{The
error bar is $6 \, 10^{-4}$ for S\&P stocks, much smaller than the observed bias except for the smallest values 
of $\theta$.} in agreement with previous studies, showing that analysts are on average over-optimistic about the 
result of companies. The bias steadily decreases as $\theta$ decreases; however, expressed in relative terms, the 
bias on the forecast {\sc eps} is still on the order of $10 \%$ one month before the {\sc ead}! Fig. 1 (right) shows how
on average the bias for a fixed $\theta$ (3 months or 6 months), $B(\theta;t)$ depends on the year of the prediction. 
One sees that the bias was very strong in the early 90's, nearly disappeared in 1993-95 to grow again during the Internet
bubble, and is back to small values since 2003 (at least for $\theta=3$ months; the bias 6 months ahead is still
significant). The reasons for a persistent positive bias in forecasts have been discussed in the past -- these can
be institutional (e.g.: analysts are paid by institutions that benefit from bullish stock markets), behavioral (e.g: 
positive recommendations are always easier to publish than negative statements) or affected by positively biased 
informations released by companies themselves.  The decrease of bias is expected because more and more 
reliable information is available as the {\sc ead} is approached. If the bias is of institutional/behavioral origin, 
the decrease is also expected since it becomes more and more untenable to publish knowingly unreliable estimates very
shortly before the true number becomes available (similar results can be observed on the UK and JP pools, with biases slightly larger than for the US one). 

From the data, we can also study how the average bias depends on the size (market value) of companies. As shown in
Fig. 2 (left), there is a clear size effect: the bias is smaller on larger size companies, probably because more people 
are concerned by the results of these companies; the pressure on analysts to make good forecasts is thus stronger. 
The size difference might then explain why the bias is larger in EU than 
for S\&P stocks; this explanation is however different
from the one proposed in \cite{Bagella} where the difference in bias is attributed to different regulations in 
the US and in the Eurozone.

We have finally studied how the bias depends on the industry sector in the US. The result is shown in Fig. 2 (see 
Table I for the sector code), which reveals that there is significant dependence of the bias on the sector. Such a
sectorization of errors will be further discussed in Section 5.

\begin{center}
\begin{tabular}{|c|l|}
\multicolumn{2}{c}{Table 1: Economics Sectors}\\
\hline 1 & Basic Materials\\
\hline 2 & Consumer, Cyclical\\
\hline 3 & Consumer, Non cyclical\\
\hline 4 & Communications\\
\hline 5 & Energy\\
\hline 6 & Finance\\
\hline 7 & Industrial\\
\hline 8 & Technology\\
\hline 9 & Utilities\\
\hline 
\end{tabular}
\end{center}

\begin{figure}
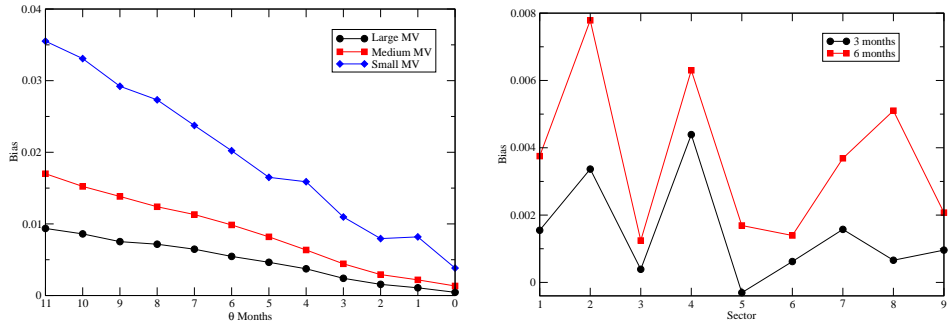

\begin{center}
\epsfig{file=biasmv.eps,width=6cm,clip=}\hspace{0.5cm}\epsfig{file=biass.eps,width=6cm,clip=} 
\end{center}
\caption{Left: The grand average bias $B(\theta)$ for US stocks, for different market values, 
as a function of the time $\theta$ to {\sc ead}. Right: The bias for S\&P stocks, for $\theta=3$ months and 
$\theta=6$ months, as a function of economic sector.}
\label{Fig2}
\end{figure}

\subsection{Forecast errors}

We now turn to the forecast error, which measures how far off, typically, the forecast deviates from the actual 
{\sc eps}. The result is shown in Figs. 3 and 4, where one sees, respectively, $\Sigma(\theta)$ as a function 
of $\theta$ and $\Sigma(\theta;t)$ as a function of $t$ for $\theta=3$ months and $\theta=6$ months. One sees that 
(i) the forecast error, much like the average bias, is larger on the global US pool than on the restricted S\&P pool, 
with the EU pool in-between;
(ii) the forecast error weakly decreases with the distance to {\sc ead}, but the final ($\theta=0$) prediction 
error is still $100 \%$ to $200 \%$ of the predicted {\sc eps} -- in other words, the forecast is not only systematically
biased, it is also very imprecise; (iii) the forecast error seems to have trended down over the years: it has on averaged 
decreased by a factor 5 between the late 80's and nowadays, with a low in the mid-nineties. 

\begin{figure}
\begin{center}
\epsfig{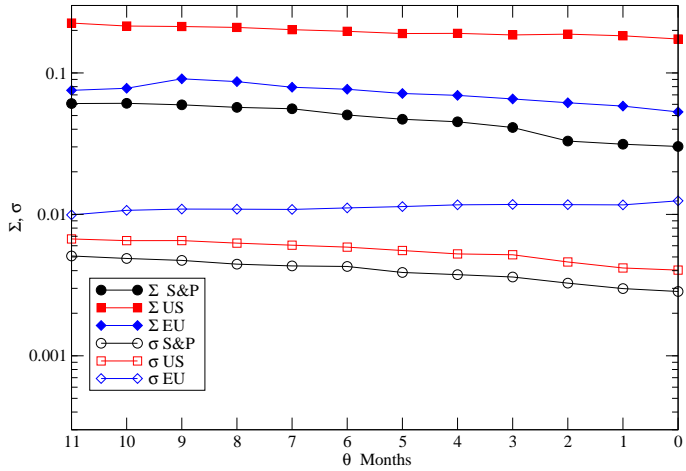}
\end{center}
\caption{The forecast error $\Sigma(\theta)$ and average dispersion $\sigma(\theta)$ for S\&P stocks, US stocks (including
S\&P stocks) and EU stocks. The forecast error is smallest for S\&P stocks and strongest for US stocks. The striking 
finding is that the forecast dispersion is on average 3-10 times smaller than the forecast error itself!}
\label{Fig3}
\end{figure}

Point (i) emphasizes the size effect observed in Fig. 2: analysts dealing with larger stocks with a wide investor base 
produce on average better forecasts than on small caps. Although intuitive, this result is in contradiction with the results of 
Bagella et al. \cite{Bagella}, who report an inverse size effect, but on a smaller sample of US stocks.\footnote{The 
argument put forward is that {\it ``it is more difficult to take into account the interaction of all performance drivers for 
a large firm with a diversified portfolio of products''}. This is however in contradiction with the usual argument 
of diversification, and actually with empirical data on the size effect on firms' growth rate dispersion \cite{Sutton}.}
We have therefore studied the restricted S\&P sample and looked for an inverted size effect there but with no success. 
In fact, small, medium and large caps of the S\&P sample have a very comparable forecast error. Hence, we 
have not been able to reproduce the inverse size effect reported in \cite{Bagella}. The difference might 
come from the selected pool of stocks, or from the treatment of errors, outliers, etc.

\begin{figure}
\begin{center}
\epsfig{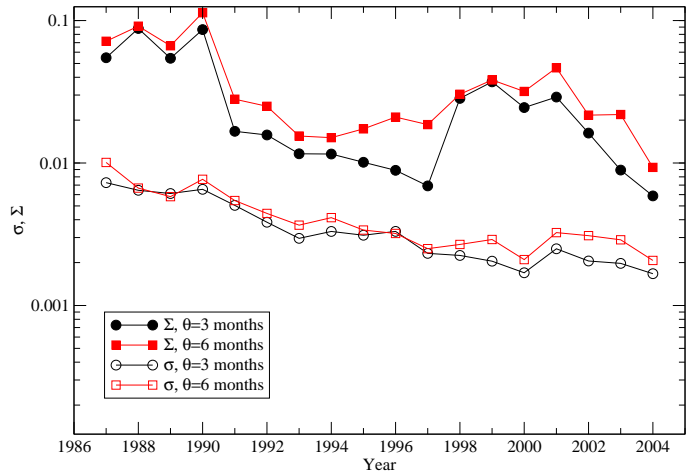} 
\end{center}
\caption{The forecast error $\Sigma(\theta;t)$ and average dispersion $\sigma(\theta;t)$ for S\&P stocks, 
for $\theta=3$ months and $\theta=6$ months, as a function of the time $t$ (year of the earning announcement). 
Both decrease with years, but the latter is always five to ten times smaller than the former.}
\label{Fig4}
\end{figure}

It is interesting to compare the performance of analysts forecasts with the simplest `no-change' forecast, which 
amounts to taking last year's {\sc eps} as an estimator of the future {\sc eps}. When compared with 
forecasts of similar maturity (i.e. choosing $\theta=11$ months), we find (see Fig. 5) that both fare very similarly 
in terms of
prediction error, but that the no-change forecast is significantly less biased. The situation improves somewhat 
for analysts as $\theta$ decreases, as of course expected.

\begin{figure}
\begin{center}
\epsfig{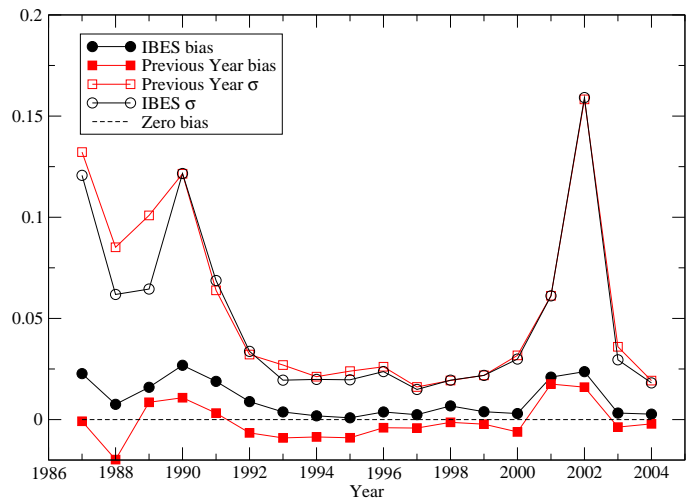} 
\end{center}
\caption{The forecast bias $B(\theta=11;t)$ and forecast error $\Sigma(\theta=11;t)$ compared to the 
same quantity computed using last year's {\sc eps} as a predictor of the future {\sc eps}.}
\label{Fig5}
\end{figure}

\section{Herding effects}

The most striking content of Figs. 3 and 4 is the fact that the forecast dispersion $\sigma$, which 
measures how much analysts disagree with each other, is {\it much smaller} than the forecast error itself 
(we find the same result for the UK and JP pools). Furthermore, the diversity of opinions decreases with time to {\sc ead}. 
These two findings strongly suggest 
herding effects, as has been discussed and modeled in several previous publications \cite{Trueman,Krause}.
One could argue that a convergence of prediction is in fact natural because all analysts are perfectly competent and 
all base their judgment on the same available information, that becomes more precise as the {\sc ead} is 
approached. This is however incompatible with the huge amplitude of the ex-post prediction error, which means that 
the information was in fact quite ambiguous and not sufficient to allow a precise forecast. The rationality scenario
would imply that the dispersion is of the same order of magnitude as the prediction error itself, i.e. $\sigma \simeq
\Sigma$. The observed inequality $\sigma \ll \Sigma$, on the other hand, can only be explained by a copy-cat
mechanism, whereby each analyst progressively biases his forecast toward the average of his fellow analysts, as
documented in \cite{Trueman}. The reasons for this strong herding behavior have been discussed in terms of e.g.
career motivations (differences between experienced and younger analysts can be observed \cite{Hong}), or behavioral 
effects, illustrating Keynes' observation that it is better to be wrong with the crowd than right against the crowd. 
In that respect, the improved quality of `outlier' (or bold) forecasts has been noticed in \cite{Hong}. 

The above observation suggests to introduce the herding ratio $\phi=\Sigma/\sigma$. From the data of Figs. 3 and 4,
we conclude that herding is noticeably stronger in the US, UK and JP ($\phi \approx 10$ for S\&P stocks, 
$\phi \approx 40$ for all US stocks, and $\phi \approx 20$ for JP and UK stocks) than in the EU ($\phi \approx 7$). 
The reason why EU appears to stand out might be related to the diversity of analysts following  
European companies. Indeed, we could speculate that 
analysts from different countries and different backgrounds might be less
prone to herding behavior.

While the dispersion of opinions is much smaller than the actual prediction error, it is still interesting to 
ask if there is some correlation between the lack of agreement between analysts and the difficulty to predict the
{\sc eps}. A scatter plot of the absolute value of prediction error $|b_{\alpha,t-\theta}|$ and the 
contemporaneous dispersion $\sigma_{\alpha,t-\theta}$ is shown in Fig. 6, together with a linear regression. 
We find, as expected, a positive intercept (meaning that even in the case of perfect agreement between analysts,
they still get the result wrong) and a slope larger than unity. The correlation between the two quantities is 
found to be $0.46$, meaning that the lack of agreement between analysts still gives some indication of how wrong
their prediction is. 

\begin{figure}
\begin{center}
\epsfig{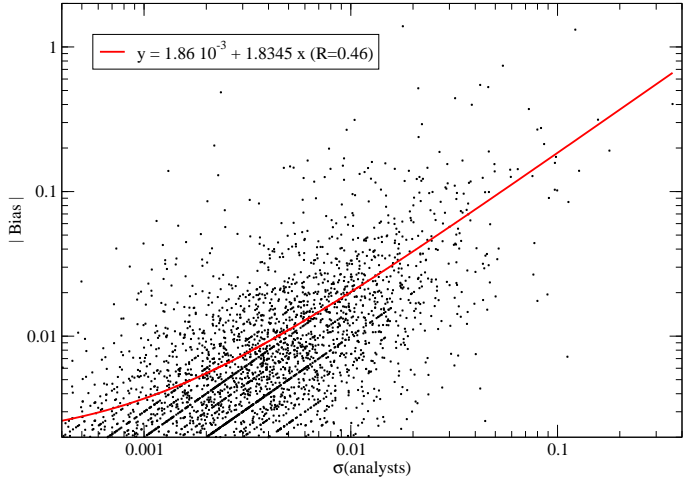} 
\end{center}
\caption{Scatter plot, for all times and all stocks, of the absolute forecast error $|b_{\alpha,t-\theta}|$ versus the 
contemporaneous dispersion $\sigma_{\alpha,t-\theta}$ for the S\&P stocks, and linear regression.}
\label{Fig6}
\end{figure}

\section{Correlation of prediction errors}

The above sections were concerned with the variance of the individual forecast errors, which we found to
be quite large. However, since we have calculated this as a grand average over all stocks, this 
large error could be localized in some sectors of activity, or some particular stocks. It is thus  
interesting to look at the correlation of these forecast errors across different economic sectors. 
One would a priori expect to see a `market mode' and a sectorization of the forecast errors, much
as stock returns. A way to investigate systematically this question is to study the covariance matrix 
$C_{\alpha,\beta}$ and the correlation matrix $c_{\alpha,\beta}$, defined as:
\be
C_{\alpha,\beta} = \langle b_{\alpha,t-\theta} b_{\beta,t-\theta} \rangle_{t,\theta} - 
\langle b_{\alpha,t-\theta} \rangle_{t,\theta} \langle b_{\beta,t-\theta} \rangle_{t,\theta},
\ee
and 
\be
c_{\alpha,\beta}=\frac{C_{\alpha,\beta}}{\sqrt{C_{\alpha,\alpha}C_{\beta,\beta}}}.
\ee
[One could also have chosen not to average over $\theta$ and have $\theta$ dependent correlation matrices].
A way to analyze the content of these matrices is to study their eigenvalues and the corresponding 
eigenvectors. We expect, much as like the correlation matrix of stock returns, that a large fraction
of these eigenvalues is dominated by measurement noise and do not contain useful information, while only the few largest eigenvalues carry meaningful information.

In order to perform this analysis, we have restricted to $M=324$ stocks from the S\&P during the period 1990-2003. 
Using results from Random Matrix Theory \cite{RMT}, we find that only three or four eigenvalues can be distinguished 
from pure noise. 
The corresponding eigenvectors are unlocalized, with an inverse participation ratio on the order of $M/2$, meaning that
a large fraction of the stocks participate to these correlation modes.\footnote{The participation ratio of a given 
normalized eigenvector $V_\alpha$ is defined as $Y=\sum_\alpha V_\alpha^4$. For a uniform eigenvector, $V_\alpha^2=1/M$ 
and $Y=1/M$, whereas for a eigenvector localized on a single stock, $Y=1$. The
quantity $1/Y$ therefore gives the number of stocks `participating' to the
eigenvector.} Theses eigenvectors corresponding to the 
largest eigenvalues of $c$ reveal that some sectors dominate, meaning that prediction errors are strong in a some sectors 
and much weaker in others. Perhaps unexpectedly, the largest eigenvector is not uniform, as one would observe if 
predictions were over-optimistic or over-pessimistic for the market as a whole. Already at the level of the largest 
eigenmodes shown on Fig. 7, some sectorization indeed appears as the largest mode is not uniform over the 
stocks; for example, the errors in sector 3 seem uncorrelated from the rest of the market. Remember that
in this sector the bias is also small on average, which is not surprising in view of the relative stability of the
demand in this sector (food, drinks, etc). The first mode also
shows that large, correlated prediction errors are localized in sectors 2, 4, 6 and 7, while
the second mode reveals clearly that sector 6 (Finance) stands out as a special sector, as far as forecast errors 
are concerned. We have also studied the $\theta$ dependent correlation matrices and found that the inverse participation ratio
of the top eigenvectors decrease as the {\sc ead} is approached, meaning that sectorization of errors increases as
$\theta$ decreases. This is again quite expected. 

\begin{figure}
\begin{center}
\epsfig{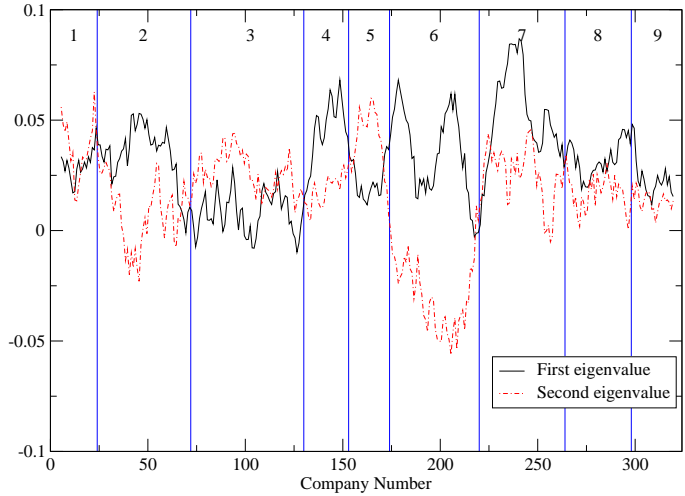} 
\end{center}
\caption{First and second eigenmode of correlation matrix $c$. Note that stocks are ordered by economics sectors. 
A smoothing of the eigenmode components has been performed for clarity.}
\label{Fig7}
\end{figure}

\section{Additional remarks and conclusion}

In this study, we have once more confirmed, on four different, large sets of data, that financial analysts are on average 
over-optimistic and show a pronounced herding behavior. These effects are however time dependent, and were particularly 
strong in the early nineties and during the Internet bubble. We have also emphasized that their forecast ability is, 
in relative terms, quite poor and comparable in quality, a year ahead, to the simplest `no change' forecast. 
As a result of herding, analysts tend to agree 
with each other five to ten times more than with the actual result. We have shown that significant differences exist between 
US stocks and EU stocks, that may partly be explained as a company size effect and not, as conjectured in \cite{Bagella}, 
to differences in regulations and information transparency. Interestingly, herding effects appear to be stronger in the 
US than in the Eurozone. Finally, we have studied the correlation of errors across stocks and shown that 
significant sectorization occurs, some sectors being easier to predict than others. 

The conclusion is that if the information available to the best experts is so scarce and/or difficult to interpret,
resulting in such imprecise and biased estimates of the earnings of companies, one can really wonder what the 
concept of efficient markets and rational pricing really means. The famous relation between the price of a stock and
the sum of expected (discounted) future dividends appears to be nearly empty with such gossamer information. In
fact, one can test quantitatively the validity of such a pricing equation. Since dividends are correlated with earnings, 
a surprise in the announced {\sc eps} compared to its anticipation should lead to an immediate change of the price of the
stock. If earnings per share and dividends per share were equal, a reasonable mean-reverting random walk model for
earnings would lead to a perfect correlation between the price change on the {\sc ead} and the difference between the
latest forecast and the actual {\sc eps}. There are however two complications: (i) dividends per share and {\sc eps}
have a correlation $\rho < 1$. For S\&P stocks, we find $\rho \approx 0.25$; (ii) on the {\sc ead}, other news 
may affect stock prices. A way to estimate their impact is to measure the average increase of volatility on {\sc ead}s
compared to normal days. For the same pool of stocks, we find a volatility ratio of $\approx 1.7$. The resulting 
correlation between stock returns on {\sc ead} and the unanticipated part of the {\sc eps} can then easily be computed to
be $\approx 0.20$, a factor four larger than the empirical correlation, that we find to be $\approx 0.05$. Such a decoupling 
between dividends and stock prices is similar to that reported by Shiller in his famous study on excess volatility \cite{Shiller}.
One could argue that the measured correlation is small because part of the information is actually already encoded in the
price, before the {\sc ead}, through `insider trading' \cite{Kyle}, or through the `collective intelligence' markets are supposed to 
be endowed with. We have checked that this is not the case by studying the correlation between price changes and {\sc eps}
changes. If the market was correctly anticipating the change of {\sc eps}, the price change between the day before 
the previous {\sc ead} and the day after the last {\sc ead} should be positively correlated with the change of {\sc eps}. 
However, we found empirically a very small, insignificant correlation of $0.02$. So the announced {\sc eps} should have 
a much stronger impact on prices, if the classical rational pricing model was correct. 

We rather believe that the notion of a true, fundamental price of a company is moot, since nobody, even the best
expert, can agree on its value. One can at best estimate a rather broad range, say within a factor of two, of
reasonable prices for a given stock \cite{Black}. Within this range, prices fluctuate freely -- arbitrage cannot 
be efficient, because of the lack of a reliable estimate of what the true price should be \cite{Shleifer}. The volatility 
is then set by the trading activity itself \cite{Odean,Hopman,QF}, and can be indeed much larger than what should be 
expected on the basis of 
the efficient market theory \cite{Shiller}. Non rational, self-referential and behavioral effects do emerge 
\cite{Kirman,cascades,Shleifer,Orlean,MG,Wyart}, 
unleashed by the lack of reliable information, generically unavailable in complex systems -- complexity puts {\it de facto} strong limits 
on the very notion of rationality \cite{Arthur,Kirman2,Galluccio}. However, as emphasized in \cite{Wyart}, on time scales such that these random 
fluctuations become of the order of 100 \% and reach the boundary of the fuzzy price range alluded to above, one 
should expect mean reversion effects to become noticeable. 
For a typical stock with a daily volatility of $3 \%$, this corresponds to 1000 days (because $100 \approx 3 \sqrt{1000}$), 
or four years. Such a time scale is precisely the typical reversion time scale discovered by De Bondt and
Thaler in their paper on overreaction in stock markets \cite{Thaler}.

Finally, the strong herding effects between analysts leaves little doubt that similar herding effects exist between
investors and market agents at large \cite{Kirman,cascades}. This is another nail in the coffin of Efficient Markets. 
Efficient Markets should 
reflect the aggregate opinion of incompletely informed, but independent agents, and therefore produce a global, unbiased 
estimator of the true value of companies. Instead, herding effects can lead to overreactions, bubbles and 
persistent mispricings, a conclusion of little surprise to the lay-man not acquainted with neo-classical economics.

\vskip 1cm
{\it Acknowledgments: }We acknowledge important discussions with J. 
Kockelkoren, L. Laloux, Y. Lemperiere and M. Potters. We also are grateful
to Thomson Financial 
for providing the data analyzed here.


\end{document}